\documentclass{osa-article}

\journal{boe}

\articletype{Research Article}

\begin{document}

\title{Highly-stable, multi-megahertz circular-ranging optical coherence tomography at 1.3 $\upmu$m}

\author{Norman Lippok,\authormark{1,2,*} Brett E. Bouma,\authormark{1,2,3} and Benjamin J. Vakoc\authormark{1,2,3}}

\address{\authormark{1}Harvard Medical School, Boston, MA 02115, USA\\
\authormark{2}Wellman Center for Photomedicine, Massachusetts General Hospital, Boston, MA 02114, USA\\
\authormark{3}Institute for Medical Engineering and Science, Massachusetts Institute of Technology, Cambridge, MA 02139, USA}

\email{\authormark{*}nlippok@mgh.harvard.edu}

\begin{abstract}
In Fourier-domain optical coherence tomography (OCT), the finite bandwidth of the acquisition electronics constrains the depth range and speed of the system. Circular-ranging (CR) OCT methods use optical-domain compression to surpass this limit. However, the CR-OCT system architectures of prior reports were limited by poor stability and were confined to the 1.55~$\upmu$m wavelength range. In this work, we describe a novel CR-OCT architecture that is free from these limitations. To ensure stable operation, temperature sensitive optical modules within the system were replaced; the kilometer-length fiber spools used in the stretched-pulse mode-locked (SPML) laser was eliminated in favor of a single 10 meter, continuously chirped fiber Bragg grating, and the interferometer's passive optical quadrature demodulation circuit was replaced by an active technique using a lithium niobate phase modulator. For improved imaging penetration in biological tissues, the system operating wavelength was shifted to a center wavelength of 1.29 $\upmu$m by leveraging the wavelength flexibility intrinsic to CFBG-based dispersive fibers. These improvements were achieved while maintaining a broad (100 nm) optical bandwidth, a long 4 cm imaging range, and a high 7.6 MHz A-line rate. By enhancing stability, simplifying overall system design, and operating at 1.3 $\upmu$m, this CR-OCT architecture will allow a broader exploration of CR-OCT in both medical and non-medical applications. 
\end{abstract}

\section{Introduction}
The imaging performance of an optical coherence tomography (OCT) system is primarily associated with the specifications of its optical subsystem (e.g., wavelength-swept laser), but also depends critically on the electronic subsystem that serves to detect, digitize, transfer, and process the output optical signals. The bandwidth of this system, for example, defines the rate at which signal information can be captured. This information capture rate directly affects a number of imaging parameters. In Fourier-domain systems such as swept-source OCT/optical frequency domain imaging or spectral-domain OCT, there is a linear relationship between the electronic bandwidth and the product of delay range and depth-scan (A-line) speed. This relationship has become increasingly relevant over the past few years, and now routinely requires a compromise of imaging parameters ~\cite{Zhao,Pfeiffer}. Circular-ranging (CR)-OCT~\cite{Siddiqui-1,Siddiqui-2,Lippok} was designed to decouple the electronic bandwidth from the image range and thereby avoid the need to compromise speed when imaging over long ranges. This decoupling is achieved by implementing an optical-domain compression that overlaps the signals from multiple, equally spaced delay locations. The first high-speed CR-OCT system was described in~\cite{Siddiqui-2}, which also presents the theory and operating principles.   

In this prior report of CR-OCT~\cite{Siddiqui-2}, the system architecture combined several innovations: a stretched-pulse frequency comb source, a passive optical quadrature demodulation circuit, and CR-specific signal processing/image processing methods. Not surprisingly, this first architecture featured a number of practical limitations that did not affect the validation of the core principles but that would, if unsolved, present meaningful barriers to continued CR-OCT development and translation. For example, the SPML laser source used portions of SMF28 and dispersion compensating fibers to generate the necessary magnitude of positive and negative dispersion. This resulted in a 45 km laser cavity that was highly sensitive to environmental temperature. As a result, it was necessary to adjust the frequency of the laser drive signals multiple times per hour. Because the system was fully phase-locked, these changes required updating downstream clocks that controlled beam scanning and signal digitization. A second source of instability was the temperature dependence of the optical quadrature demodulation circuit. This was overcome by periodic re-calibration of the circuit~\cite{Siddiqui-3}. Although effective, the recalibration protocols were tedious and served to slow the pace of development. A third deficiency of the prior design, unrelated to stability, was its inability to operate away from 1.55 $\upmu$m (imposed by the dispersion of the SMF28 and dispersion compensating fibers).  

We present in this work an improved CR-OCT system architecture that addresses the instability of the prior design and extends imaging to the 1.3 $\upmu$m range. We demonstrate for the first time CR-OCT imaging based on a CFBG-SPML laser. By using a CFBG for dispersion, the laser stability was dramatically improved and the laser could be operated at 1.3 $\upmu$m. We implemented a new, highly-stable method for optical in-phase (I) and quadrature (Q) signal generation based on active phase-shifting using a LiNbO$_3$ electro-optic modulator (EOM); two I/Q generation protocols are described, validated, and demonstrated. CR-OCT imaging using these innovations and operating at imaging speeds of 7.6 MHz with a optical bandwidth from 1240-1340 nm and a coherence lengths of 4 cm is presented. To date, this work describes the most compact and stable high speed CR-OCT system and can serve to accelerate the exploration of the architecture in a broad range of imaging applications. 

\section{Experimental setup}

\begin{figure}[t]
\centering{\includegraphics[width=\textwidth]{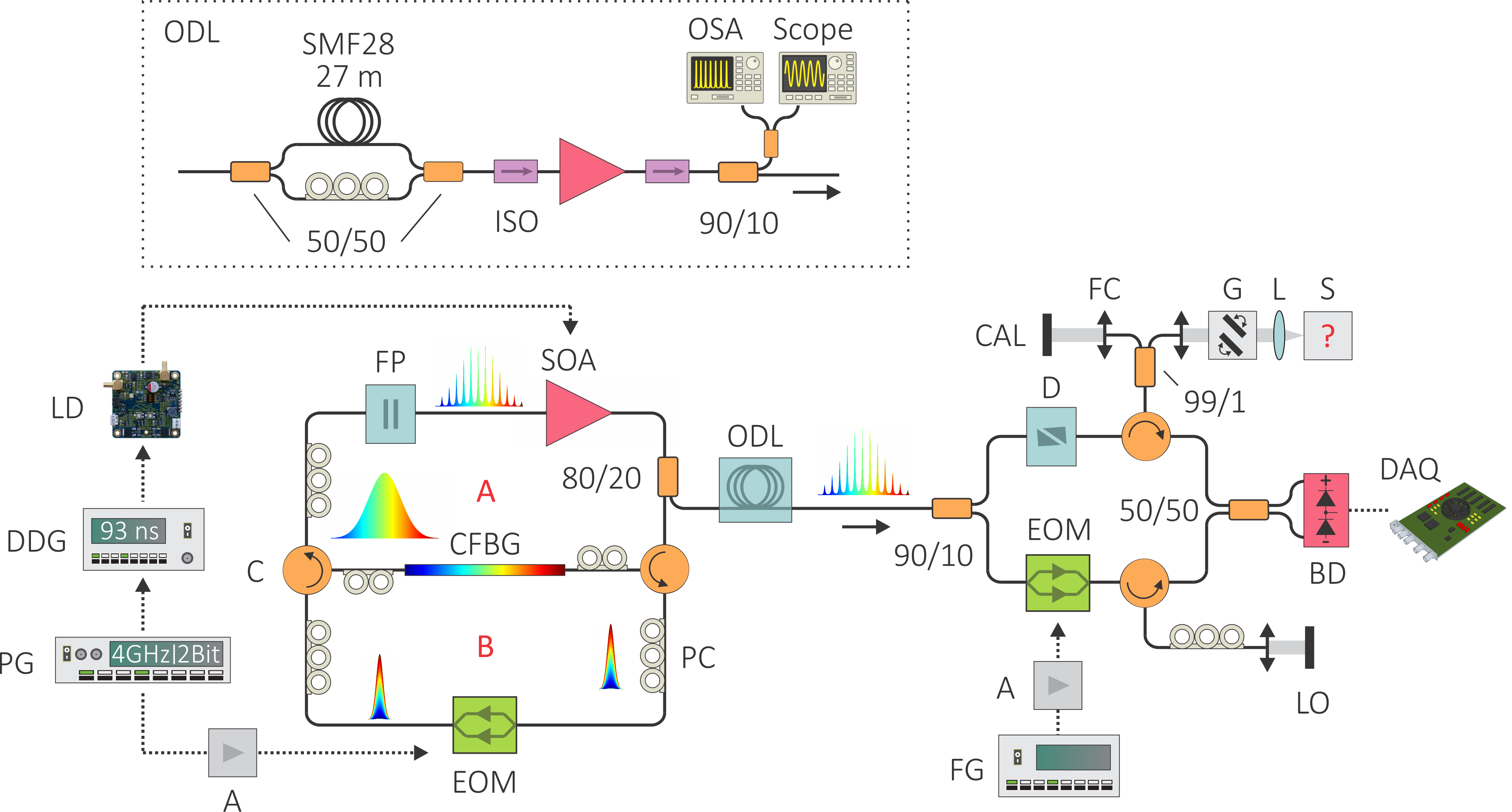}}
\caption{Experimental setup showing the frequency comb SPML laser and circular ranging OCT system: LD, laser diode driver; DDG, digital delay generator; PG, pattern generator; A, amplifier; EOM, electro-optical modulator; PC, polarization controller; CFBG, continuous fiber Bragg grating; SOA, semiconductor optical amplifier; FP, Fabry-P\'erot etalon spectral filter; ODL, optical delay line; OSA, optical spectrum analyzer; ISO, optical isolator; FG, signal (function) generator; D, dispersion compensation; LO, local oscillator; CAL, calibration signal; FC, fiber collimator; G, galvanometer mirrors; L, lens; S, sample; BD, balanced photodiode; DAQ, data acquisition board.}
\label{Fig1}
\end{figure}

Figure~\ref{Fig1} shows the CR-OCT system configuration. The laser was based on a theta-cavity SPML laser design using a CFBG, and additional details of its operation are provided in~\cite{Khaz}. An electro-optic modulator (EOM) (MX1300-LN-10, Photline) was driven at a harmonic of the cavity round-trip time (3.8 MHz). The drive pulse for the EOM was generated by a pattern generator (PAT 5000, SYMPULS) followed by an RF amplifier (DR-PL-10-MO, iXBlue). The driving pulse width was adjusted to 520 ps. A signal generator (SG384, Stanford Research Systems, Inc.) was used to generate an external clock signal for the pattern generator. A continuous fiber Bragg grating (CFBG, Proximion) was placed between two circulators to access both normal and anomalous dispersion. The ~9.5 m grating was designed to produce a linear group delay with respect to optical frequency over a continuous wavelength range from 1240 nm to 1340 nm ($\Delta\lambda=$100 nm). The dispersion at 1290 nm was $\pm$930 ps/nm, and the group delay variation across the full bandwidth of the CFBG was 93 ns. Because the same CFBG was used for both positive and negative dispersion, a high degree of dispersion matching was achieved. The laser output was taken after the SOA using a 20\% output tap coupler. 

To generate a frequency-stepped output, an air-spaced Fabry-P\'erot (FP) etalon was included in the cavity. The etalon was constructed using two dielectric mirrors (Korea Electro-Optics Co., LTD.) with a reflectivity of 85\% on one side and an wedged opposite surface. The measured single-pass finesse was 12 and the free spectral range (FSR) was adjustable through translation of one of the mirrors. We note that the measured single pass finesse was lower than the theoretical finesse of 18, likely due to surface irregularity and/or deviations from the target reflectivity. This FP etalon frequency comb was \emph{nested} on top of the $\sim$3.8 MHz frequency comb of the actively mode-locked laser cavity, and only the \emph{nested} frequency comb defines relevant imaging parameters (i.e., circular depth range). Within the CFBG passband, approximately 30\% of the light was transmitted through the CFBG, creating three optical cavities (A, B, AB). To suppress light circulating in cavities A and B, we used SOA modulation (T160, Highland Technology) at a frequency given by the cavity AB roundtrip time, with an on-state determined by the CFBG dispersion (93 ns). The modulation was applied using a digital delay generator (DG645, Stanford Research Systems) triggered by the pattern generator. The external clock generator, pattern generator and digital delay generator were phase synchronized using a 10 MHz reference signal with the clock generator acting as the master clock. The source duty cycle was 38\%, which yielded a 3.8 MHz repetition rate at the laser output. The repetition rate was doubled to 7.6 MHz (76\% duty cycle) using a 27~m SMF28 fiber delay line. Finally, a booster amplifier was included after the delay lines to increase the average output power to 80 mW.

The interferometer reference arm included a LiNbO$_3$ electro-optic phase modulator (EO Space) that was used to generate in-phase (I) and quadrature (Q) fringe signals that together comprise the complex fringe signal needed in CR. The EOM was designed for a wavelength region at 1.3 $\upmu$m and was made of a polarizing waveguide (no integrated polarizer). The RF bandwidth was 10 GHz, the insertion loss was 3 dB, and $\pi$-voltage of 5.3~V. For modulations above 30 kHz, the electrical signal was amplified using a broadband amplifier (MTC5515, MultiLink Tech Corp.). In the sample arm, a galvanometer (504 Hz, Thorlabs or 4 kHz, EOPC) enabled two-dimensional scanning (more details in Sec.~\ref{fdemod}, \ref{ademod}). Imaging was performed using a lens with focal length of 50 mm that offered a spot size of 80 $\upmu$m. Dispersion matching in the sample arm accounted for waveguide dispersion from the EOM. Signals were acquired using a 1.6 GHz balanced detector and a 4 GS/s, 12 bit data acquisition board (AlazarTech, ATS9373).

\begin{figure}[t]
\centering{\includegraphics[width=\textwidth]{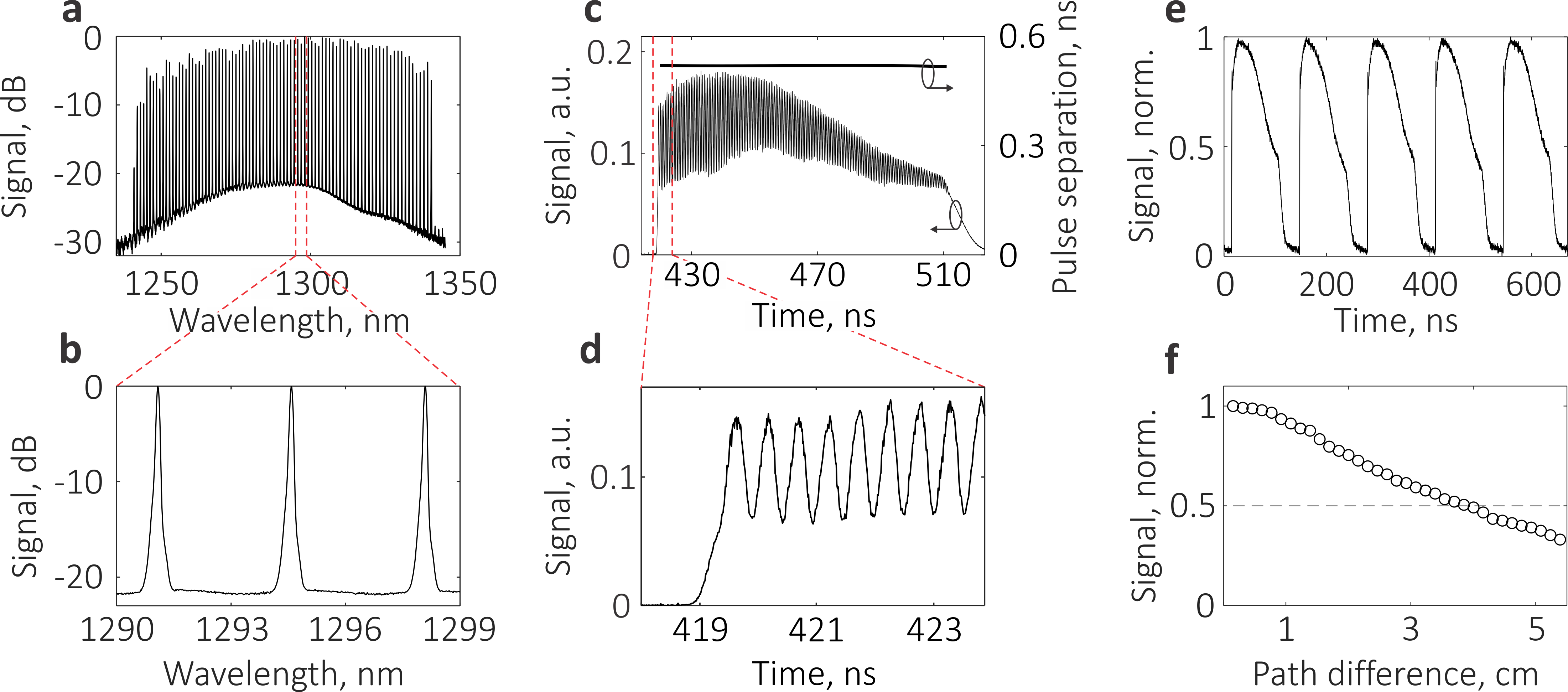}}
\caption{Measured SPML laser performance. (a) Optical spectrum. (b) Magnified optical spectrum. (c) Time trace of a single sweep using a 260 ps EOM drive (electrical) pulse. (d) Magnified time trace showing pulsation. (e) Time trace showing five A-lines. (f) Measured fringe amplitude versus path difference.}
\label{Fig2}
\end{figure}

\section{1.3 $\upmu$m CFBG-SPML performance}

The CFBG-SPML laser performance is presented in Figure 2. The laser generated a frequency comb output matching to the FP etalon [Fig.~\ref{Fig2}(a,b)]. We confirmed that the laser output followed the etalon FSR within the tested range of 50 to 450 GHz. With the FSR set to 100 GHz and a 260 ps EOM drive pulse, the laser generated the anticipated pulse train [Fig.~\ref{Fig2}(c,d)]. The time-domain output was recorded using a 35 GHz photo detector (1474-A, New Focus) and a 20 GHz sampling scope (HP 54120B). In Fig.~\ref{Fig2}(c), the pulse separation was measured to be 0.52 +/- 0.03 ns across the operating bandwidth. This is in excellent agreement with the predicted pulse repetition rate given by $f_p=D/$\emph{fsr}, where $D$ is the dispersion of the CFBG  (930 ps/nm or 194~THz/$\upmu$s). Note that an electrical pulse of 260 ps was used in Fig.~\ref{Fig2}(c,d) to allow the discrete optical pulses to be resolved in the time-domain. The remaining panels of Fig.~\ref{Fig2} used a longer 520 ps electrical pulse. In Fig.~\ref{Fig2}(c,d), the measured optical pulse width ranged between 300 and 470~ps across the lasing bandwidth, broader than the 260~ps EOM drive pulse. This is likely due to the interplay between the nonlinear broadening in the SOA and dispersion within the FP etalon passband. A train of five sweeps is shown in Fig.~\ref{Fig2}(e) demonstrating a 7.6~MHz repetition rate with a 76\% duty cycle. The trace was recorded using a 520 ps electrical pulse at the EOM and a 2 GHz real-time oscilloscope (Tektronix, MSO5204). Pulsation was not observed in this case due to a higher optical pulse duty cycle and a lower RF detection bandwidth. The coherence length of the laser after post-amplification was measured to be 4 cm [6 dB roll-off, Fig.~\ref{Fig2}(f)]. 

\section{Active Methods for I/Q Signal Generation}
Numerous methods to resolve depth degeneracy were implemented in conventional OCT~\cite{Hofer,Choma,Vakoc,Wojtkowski,Leitgeb-1,Yun,Davis,Zhang,Tao,Yasuno,Leitgeb-2,Baumann,Vakhtin}. In prior CR-OCT demonstrations, a passive method with poor stability was employed~\cite{Siddiqui-3}. We recently demonstrated active quadrature detection by frequency shifting, using an acousto-optic-modulator~\cite{Lippok}. While easily implemented, this method was limited to A-line rates of only a few megahertz. In this work, we implemented two protocols for generating complex (I and Q) fringe signals using a LiNbO$_3$ electro-optic phase modulator within the reference arm. The first modulates between imaging frames (inter-frame I/Q generation), while the second modulates between A-lines (inter-Aline I/Q generation). Each of these protocols was shown to dramatically improve stability. The performance and operation of each of these protocols are described in this section, followed by a discussion on the relative advantages and limitations of each.  

\subsection{Inter-frame I/Q signal generation}\label{fdemod}

\begin{figure}[t]
\centering{\includegraphics[width=10.2cm]{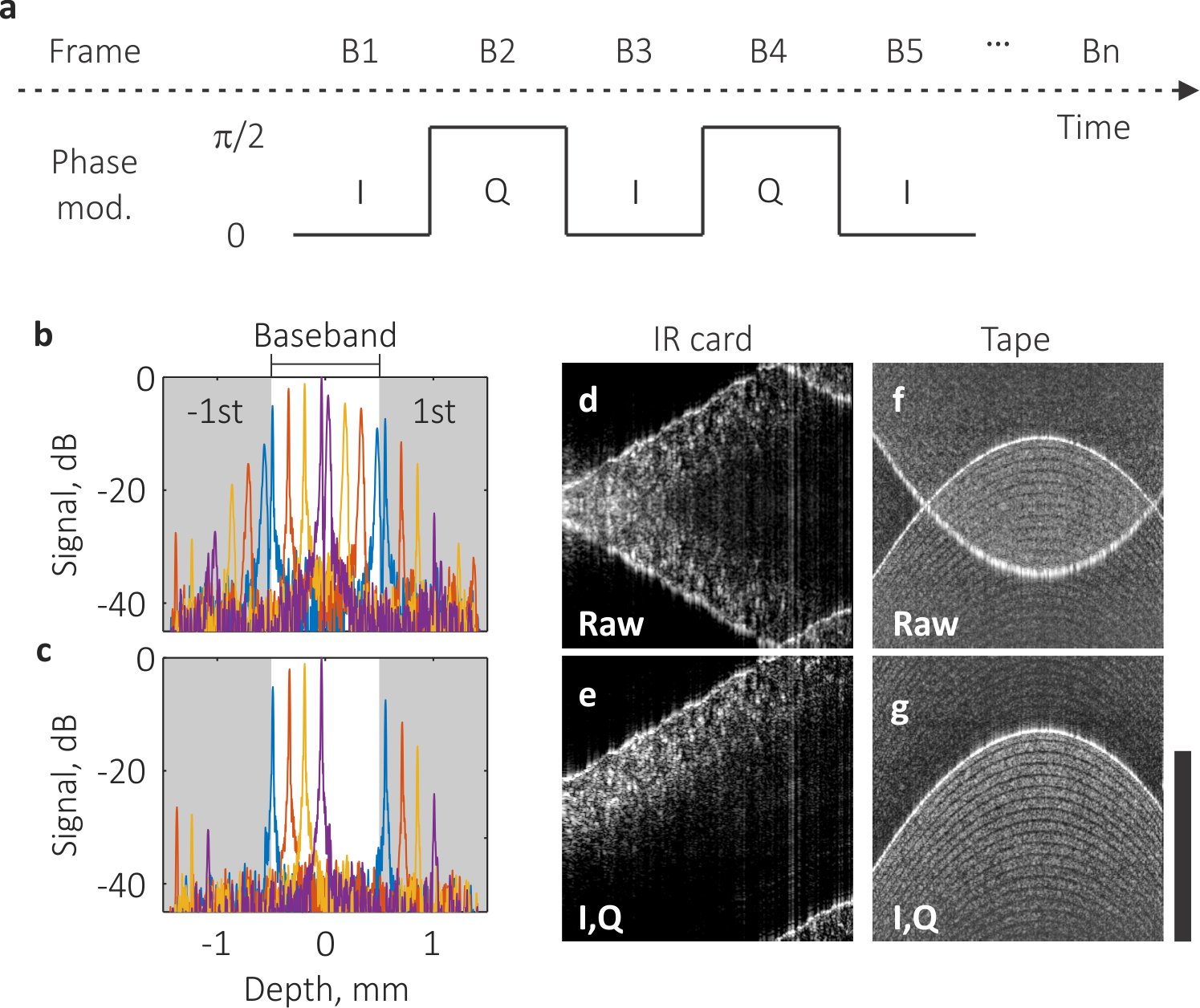}}
\caption{Frame demodulation. (a) Schematic of frame demodulation. (b),(c) Artefact suppression for mirror signals across the baseband showing the raw, dispersion compensated signals (b) and I/Q signals (c). FSR = 150 GHz. The -1st, baseband and 1st order signals are indicated. (d),(e) Imaging of an IR card showing the raw, dispersion compensated image (d) and I,Q demodulated image (e). (f),(g) Imaging of adhesive tape showing the raw, dispersion compensated image (f) and I,Q demodulated image (g). A FSR of 100~GHz was used during imaging. Scale bars correspond to 1 mm.}
\label{Fig3}
\end{figure}

We implemented a straightforward frame-based I/Q signal generation approach that acquires full frames at a constant phase, with discrete phase shifts of $\pi/2$ between frames [Fig. 3(a)]. This results in sequential acquisition of I and Q frames, which are combined in post-processing to create a single complex (I + $i$Q) frame. 

Beam scanning was performed by a resonant scanner operating at 3908.17 Hz (PLD-1S, EOPC). We note that the resonant scanner frequency was slightly detuned from its mechanical resonance to match an integer of the laser repetition rate for synchronization, providing exactly 1932 A-lines per scanning cycle. The phase modulation frequency was adjusted to half the frame rate, $f_{\mathrm{PM}}=$ 1.95~kHz. Signals were digitized at a clock rate of $f_{\mathrm{s}}=$ 3.87~GS/s, which provided multiple samples per optical pulse. 

Figures 3(b,c) present measured point spread functions (PSFs) across the circular depth range using only the I frames [Fig. 3(b)] or using the complex I/Q frames [Fig. 3(c)]. The suppression of the complex conjugate signals in the I/Q frames relative to that of the I frames  exceeded 40~dB (limited by measurement SNR). Imaging was performed using a FP with a 150~GHz FSR (circular depth range of 1 mm) [Fig. 3(d-g)]. The images present data within the circular depth range only as generated using either the I frames [Fig. 3(d,f)] or using the complex (I+$i$Q) frames [Fig. 3(e,g)]. Here, the removal of complex conjugate signals can be clearly appreciated. Note that the desired circular wrapping of the sample signals can be observed [Fig. 3(e)]. While this demonstration used only A-lines collected during forward scanning, it may be possible to modulate between scan directions to double the acquisition speed. 

\begin{figure}[t]
\centering{\includegraphics[width=\textwidth]{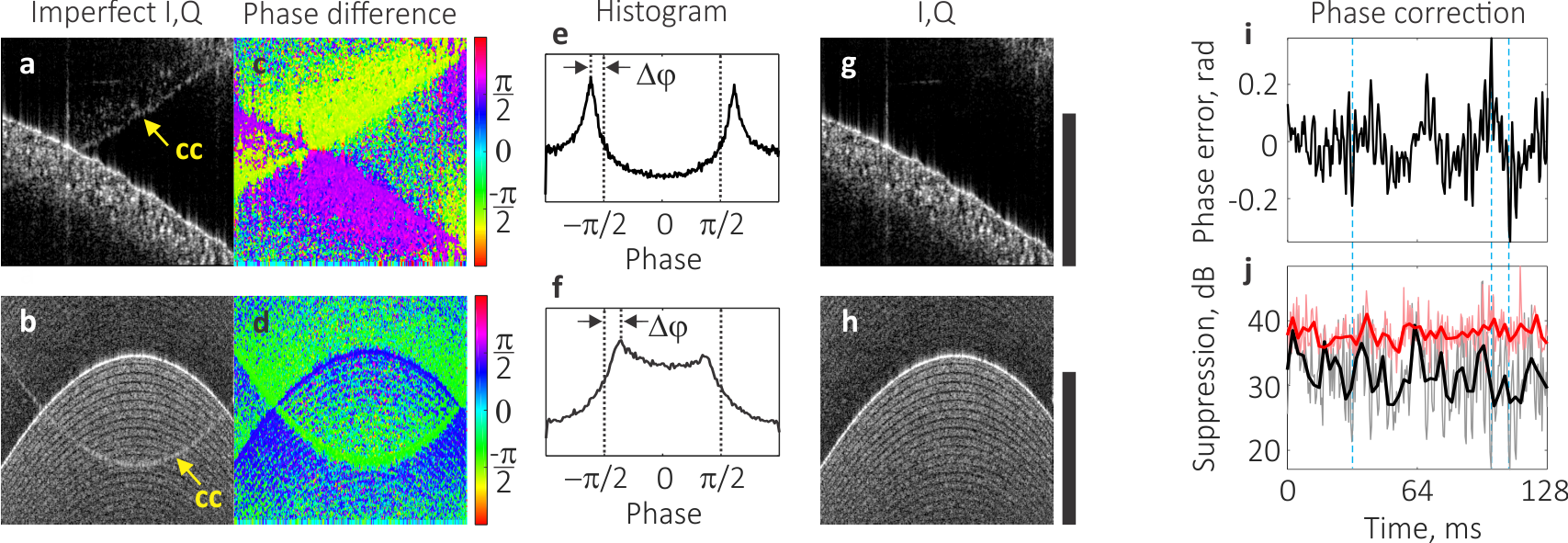}}
\caption{Illustration of phase correction using an IR card (top row) and adhesive tape (bottom row). (a),(b) Images with remaining artefacts after I,Q demodulation (yellow arrow). (c),(d) Phase difference between phase modulated frames, $\varphi(x,z)=\arg\{S_I(x,z)S_Q^*(x,z)\}$. (e),(f) Phase histogram showing $\varphi$ occurrences from data shown in (c),(d). The histogram was used to obtain a global phase offset, $\Delta\varphi$, from the quadrature point. (g),(h) Phase corrected images. (i) Phase error (offset) from the ideal quadrature point for 251 continuously recorded frames. (j) Corresponding measured suppression due to the phase error in (i) before (black line) and after phase correction (red line). The thick lines show the averaged suppression using 5 frames. Scale bars correspond to 1 mm.}
\label{Fig4}
\end{figure}

The frame-based I/Q generation method assumes that the measured I and Q frames differ only by the differential phase shift applied to the reference arm field by the EOM. In the presence of sample motion, either Doppler phase shifts due to axial motion or stochastic phase shifts due to transverse motion can occur and degrade the quadrature relationship between frames. The magnitude of this effect is somewhat controlled by using a fast 4 kHz frame rate, but significant motion-induced phase shifts can still occur. This is demonstrated in Fig. 4(a,b) for the same samples placed on a mechanically unstable stage. 

To mitigate these motion effects on I/Q signal generation, we implemented a post-processing method to cancel sub-wavelength axial motion. The principle is straightforward: if the motion-induced phase shift is small and constant across the frame, then the phase shift induced by motion can be measured directly (as the deviation, $\Delta\varphi$, from $\pi/2$ averaged across the frame). This phase-shift can then be corrected by applying its negative to one of the measured frames and thereby recovering the quadrature relationship between frames. 

For the two acquisitions shown in Figs. 4(a,b), we first display the measured phase difference between the I and Q frames. Note that the phase data in Fig. 4(c) clusters around +/-$\pi/2$, with the sign being positive for the primary image and negative for the complex conjugate image (the definition of primary and complex conjugate are only meaningful in comparison to each other; either could be classified as primary, with the other being the conjugate). In this image, the phase-measurements deviate significantly from these  +/- $\pi/2$ values for pixels in which the signal is low, or for pixels in which there is overlap of primary and conjugate signals of approximately equal power. This later overlap noise can be seen most clearly in the more transparent object of Fig. 4(b,d). 

The phase difference histograms for each image [Fig. 4(e,f)] reveal the motion-induced shifting of the mean phase difference away from $\pi/2$. The noise (due to low SNR or overlap) broadens the distributions as seen most clearly in Fig. 4(f). We extracted the motion-induced phase error, $\Delta\varphi$, as  the difference from the negative histogram peak and -$\pi/2$ [Figs. 4(e,f)]. We then applied the negative of this phase shift to the measured Q frame before constructing the complex frame as I+$e^{i(\pi/2-\Delta\varphi)}$Q. The resulting images [Fig. 4(g,h)] show improved complex conjugate suppression relative to the uncorrected images [Fig. 4(a,b)]. We note that the computational burden of this correction is minimal as the phase signals are available directly from the Fourier-transformed fringes. Figures 4(i,j) show the motion-induced phase error, $\Delta\varphi$, and the corresponding complex conjugate suppression before and after correction for 250 frames over a time span of 128~ms. 

\begin{figure}[t]
\centering{\includegraphics[width=6.8cm]{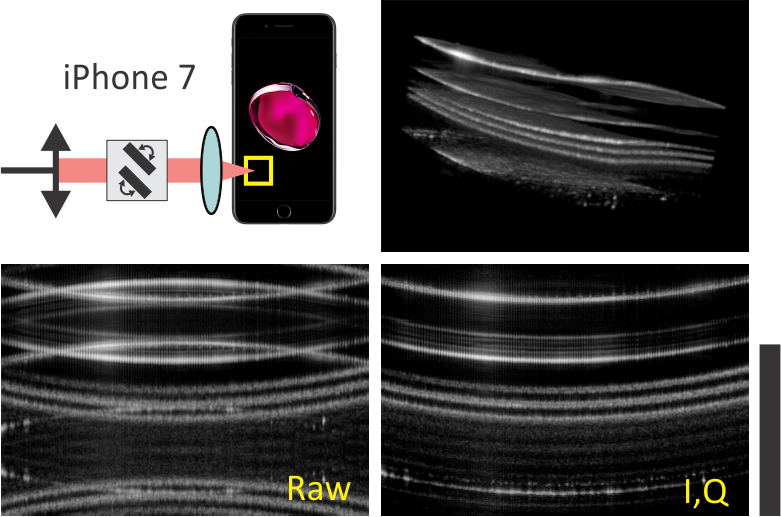}}
\caption{Imaging of an iPhone 7 display using only the I frames (left panel), or using the complex (I+$e^{i(\pi/2-\Delta\varphi)}$Q) frames with motion correction (right panel). Distinct layers beneath the top glass plate are visible. Scale bars correspond to 1 mm.}
\label{Fig5}
\end{figure}

A further imaging example of an iPhone display is presented in Fig. 5. The raw and I,Q demodulated image as well as a rendered visualization of a volumetric image is shown, clearly highlighting numerous distinct layers below the surface. This also demonstrates the strong suitability of SPML-based circular ranging by frame demodulation for industrial applications, where volumetric video-rate and long range imaging could be beneficial, including wide field-of-view material, display or paint inspection.

This technique can be used to improve I/Q performance when it is degraded by small-amplitude axial motion. It is important to recognize, however, that it will provide minimal benefit in the presence of large motion (i.e., when $\Delta\varphi$ approaches $\pi/2$), or in the presence of transverse decorrelating motion. In these scenarios, the Aline-based I/Q generation (discussed next) may be preferred. 

\subsection{Inter-Aline I/Q signal generation}\label{ademod}

\begin{figure}[t]
\centering{\includegraphics[width=11cm]{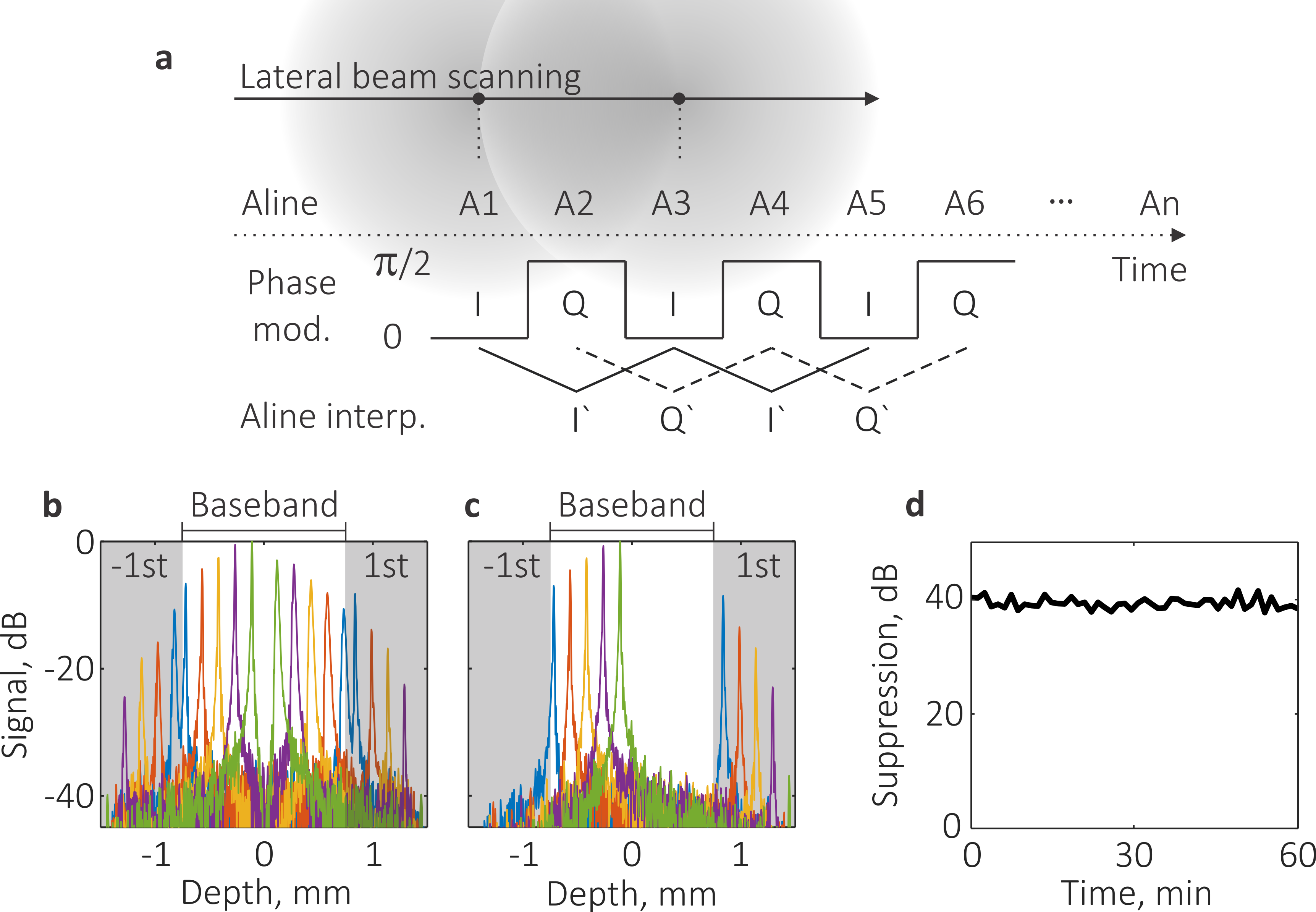}}
\caption{A-line I/Q signal generation. (a) Schematic of A-line demodulation. (b),(c) Artefact suppression for mirror signals across the baseband showing the raw, dispersion compensated signals (b) and I,Q demodulated signals (c). FSR = 100 GHz. The -1st, baseband and 1st order signals are indicated. (d) Measured complex conjugate suppression of a mirror signal over a time period of one hour.}
\label{Fig6}
\end{figure}

In A-line I/Q signal generation, the phase modulator in the reference arm induced phase shifts between alternating A-lines during lateral scanning. Nominally, this would yield alternating I and Q A-lines, and adjacent A-lines could be combined (I+$i$Q) to generate complex A-lines. However, due to beam scanning, adjacent A-lines measure slightly different sample portions, and this induced a decorrelation that produces quadrature errors. To reduce these quadrature errors, we calculated interpolated A-lines from every-other A-line, and used these interpolated A-lines when forming complex A-lines, as illustrated in Fig.~\ref{Fig6}(a). 

A galvanometric scanner (Thorlabs) was employed with a fast axis frequency of 504.3~Hz. The frequency was set to a multiple integer of the master clock (pattern generator external clock). The phase modulation frequency was adjusted to half the A-line rate, $f_{\mathrm{PM}}=$ 3.8~MHz. This had to be carefully selected to match a multiple integer of the laser repetition rate for synchronization. The sampling rate was adjusted to $f_{\mathrm{s}}=$ 3.87~GS/s, which conveniently matched the pattern generator clock rate, yielding 389 points per A-line.

Figures~\ref{Fig6}(b) and (c) show A-lines PSFs using the I A-lines [Fig.~\ref{Fig6}(b)] and complex, interpolated A-lines [Fig.~\ref{Fig6}(c)]. Note that no lateral beam scanning was applied in this case, and thus the A-line interpolation, while used, was unnecessary. The suppression of complex conjugate artefacts was approximately 40 dB, and was limited to the system noise floor, i.e., suppression was greater than 40 dB). Coherent averaging (100 A-lines) experiments (not shown) revealed that the suppression exceeded 50 dB. The I/Q signal generation was highly stable, and it was not necessary to readjust any of the drive signals for at least several days. The suppression levels measured continuously over one hour are shown in Fig.~\ref{Fig6}(d).

Figure 7(a) demonstrates the measured complex conjugate term suppression as a function of beam step size normalized to the beam diameter, $\Delta x/\delta x$, where $\delta x$ corresponds to twice the Gaussian beam waist parameter, $\delta x = 49~\upmu$m. In this work, the beam step size was adjusted by changing the scanning amplitude, while maintaining the scanning frequency. Complex interpolation enhanced suppression by approximately 10 dB for step sizes on the order of one quarter of the beam spot size or less. With moderate spatial oversampling ($\Delta x/\delta x$ = 0.2) and complex interpolation, complex conjugate suppression equal to 35 dB or better was achieved. For minimal oversampling ($\Delta x/\delta x$ = 0.4) and complex interpolation, suppression of 20 dB was achieved. For reference, we note that $\Delta x/\delta x$ = 0.5 is often used in conventional OCT imaging. Finally, we note that the 40 dB suppression limit for small step sizes was due to the system noise floor and does not represent the hard limit of this technique. Figure 7(b) shows an example of complex interpolation by imaging an IR card using a beam step size of $\Delta x/\delta x=0.23$. The raw image clearly shows complex conjugate terms (cc). I,Q demodulation offers suppression but artefacts remained visible. Complex interpolation further improves suppression, reducing the artefacts to the -30~dB noise floor that was typically observed during imaging. Finally, the demodulated and corrected image was stitched three times to make the borderless wrapping of sample structure that exceeded the circular depth range more clear.  

\begin{figure}[t]
\centering{\includegraphics[width=11cm]{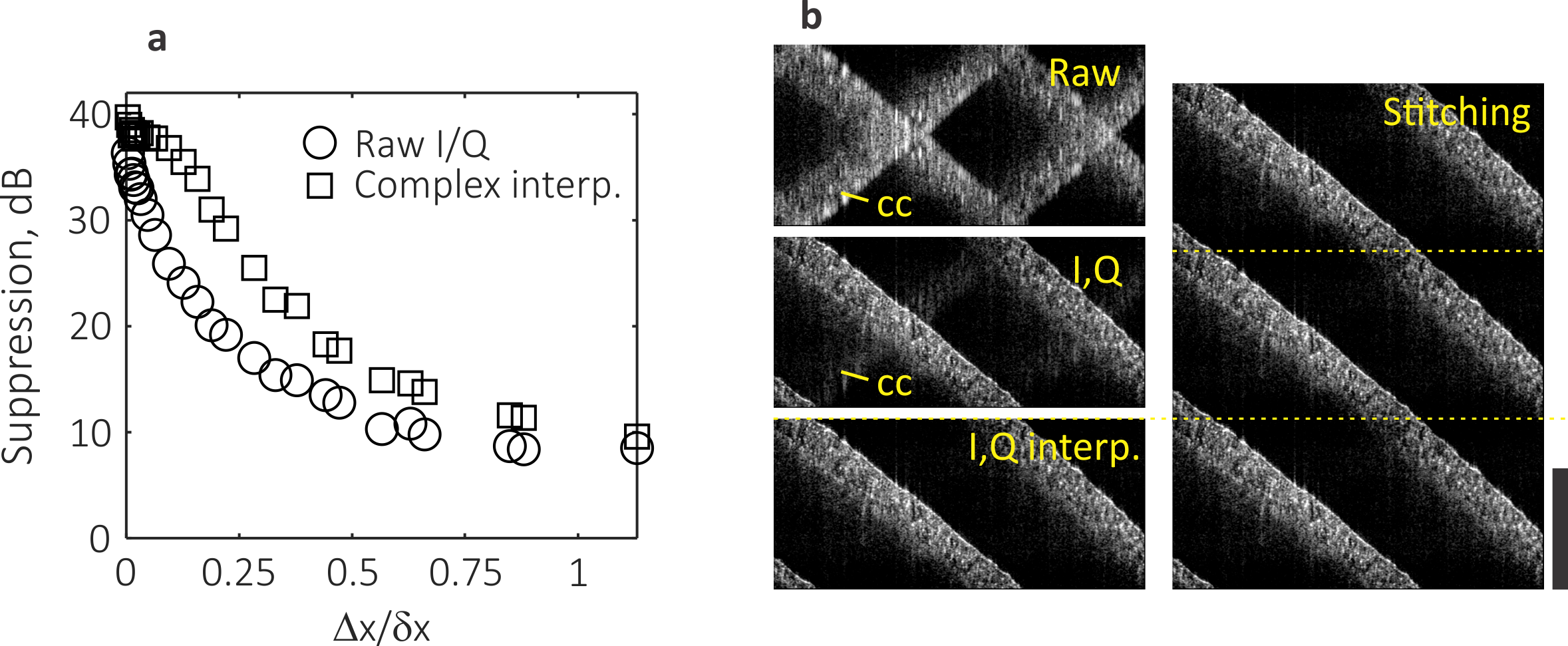}}
\caption{Complex interpolation. (a) Measured suppression before (circles) and after complex interpolation (squares) as a function of beam step size, $\Delta x$, normalized to the beam spot size, $\delta x$. (b) Imaging example of an IR card illustrating complex interpolation. It is shown the raw image and I,Q demodulated image. The demodulated image was stitched three times to make the borderless wrapping of sample structure beyond the baseband range clear. The FSR was 100 GHz ($L_B=$ 1.5 mm). Scale bar corresponds to 1 mm.}
\label{Fig7}
\end{figure}

\begin{figure}[b]
\centering{\includegraphics[width=\linewidth]{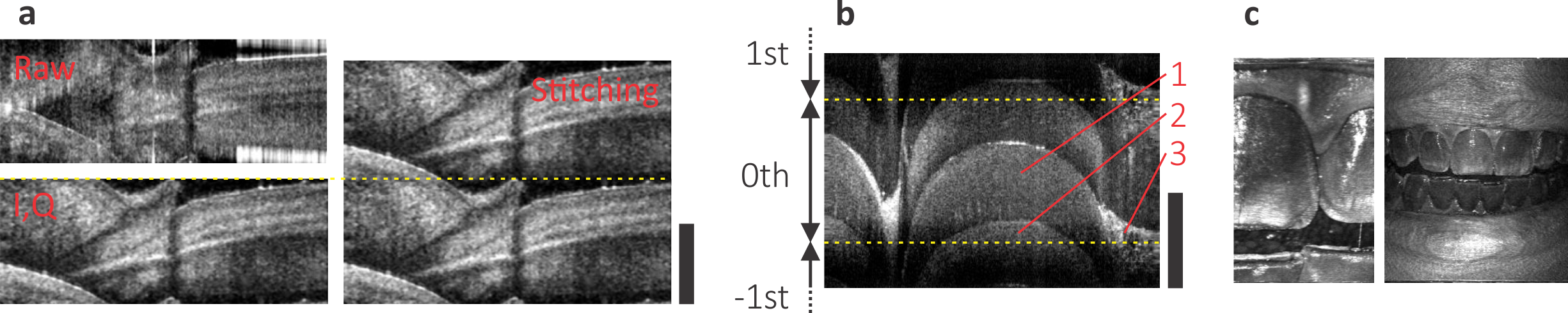}}
\caption{Imaging examples using inter-Aline I/Q signal generation. (a) Cross-sectional images of a human nail fold showing the image generated by the in-phase Alines and the image generated by the complex Alines. The demodulated image is stitched twice to reveal the the borderless wrapping of structure across the circular depth range boundary. (b) Complex image of a human tooth showing the enamel (1), dentin (2) and gum (3). (c) Volumetric depth projections showing human teeth, using a 50 mm lens (left) and 150 mm lens (right).} 
\label{Fig8}
\end{figure}

An imaging example of a human nail fold is demonstrated in Fig.~\ref{Fig8}(a). It is shown the raw, dispersion compensated image as well as the artefact free I,Q demodulated image. The beam step size was $\Delta x/\delta x=$ 0.15. No complex interpolation was applied. A FSR of 100 GHz was used and the images only show the baseband range, i.e., $L_B=$ 1.5 mm. The I,Q demodulated image was stitched twice on top of itself in order to make the borderless wrapping of sample structure exceeding the baseband more clear. Moreover, Fig.~\ref{Fig8}(b,c) shows imaging of human teeth. Distinct layers showing the enamel (1), dentin (2) and gum (3) are visible in Fig.~\ref{Fig8}(b). No complex interpolation was required. Sample structure exceeding the baseband at the top into the 1st order is folded back into the bottom of the baseband, whereas structure reaching into the -1st order at the bottom is wrapped back into the top of the baseband. The structure above the enamel surface is thus the bottom of the tooth that is folded into the top of the baseband. The high imaging depth of teeth suggests a larger baseband (smaller FSR) in order to avoid overlapping structure. Finally, SPML-based circular ranging is used to demonstrate volumetric, video camera like imaging of teeth at 15 volumes per second, using a 50 mm lens [Fig.~\ref{Fig8}(c), left] and 150 mm lens [Fig.~\ref{Fig8}(c), right]. As for the frame demodulation technique (Section~\ref{fdemod}), this represents a 50-fold compression factor in digitizer bandwidth and data load for signal and image processing compared to continuously swept laser with similar performance. Together with the imaging of layers in Fig.~\ref{Fig8}(b), this illustrates the versatility of medical and industrial applications offered by the simultaneous high speed and long range of circular ranging.

\section{Conclusion}

We have demonstrated a stable, CFBG-based frequency comb SPML laser for CR-OCT at 1.3 $\upmu$m for the first time. The laser had an A-line rate of 7.6 MHz, which could be increased to 10 MHz (given by CFBG length) by optimizing the laser cavity fiber lengths. The optical bandwidth was 100 nm and coherence length was 4 cm. The use of a CFBG enables a linear-in-wavenumber group delay that avoids the need for interpolation in post-processing. Circular ranging OCT was implemented by using an active lithium niobate modulator to generate I/Q signals used to construct complex signals. Waveguide-based lithium niobate devices provide a straightforward and highly configurable method to modulate phase, operate comfortably in the GHz range and thus pose no immediate speed limit for CR-OCT. Two I/Q protocols were described: inter-frame and inter-Aline. The inter-frame method did not impose a scanning speed limitation and offered a frame rate of 1.95 kHz (complex frame), but is susceptible to motion-induced phase-errors that deteriorate the complex signal extinction. A phase correction approach to reduce the magnitude of this effect for small magnitude axial motion was presented. The inter-Aline method is largely immune to sample motion but restricts scanning speed or scanning angle, with a demonstrated frame rate of 504 Hz. For sufficient oversampling, neighboring A-line were correlated and offered suppression greater than 40 dB.

\end{document}